\documentclass[%
 reprint,superscriptaddress,amsmath,amssymb,
 aps,nofootinbib]{revtex4-1}
\usepackage{dsfont}
\usepackage{graphicx}

\usepackage{dcolumn}
\usepackage{bm}
\usepackage{color}
\usepackage{xcolor}
\usepackage{epsfig}
\usepackage{soul}
\usepackage[colorlinks=true,urlcolor=brown,linkcolor=blue,citecolor=magenta]{hyperref}
\usepackage{natbib}
\usepackage{float}
\usepackage{footmisc}
\usepackage{siunitx}

\definecolor{lightred}{rgb}{1,0.5,0.5}
\definecolor{lightgreen}{rgb}{0.5,1,0.5}
\definecolor{lightblue}{rgb}{0.5,0.5,1}
\definecolor{lightcyan}{rgb}{0.5,0.75,0.75}
\definecolor{lightmagenta}{rgb}{0.75,0.5,0.75}
\definecolor{customgreen}{rgb}{0.494,1,0.502}

\newcommand{\GeV}{\mathinner{\mathrm{GeV}}}

\newcommand{\Pdd}{\Phi_1^\dagger\Phi_1}
\newcommand{\Puu}{\Phi_2^\dagger\Phi_2}
\newcommand{\Pdu}{\Phi_1^\dagger\Phi_2}
\newcommand{\Pud}{\Phi_2^\dagger\Phi_1}

\begin{document}

\title{Primordial Magnetogenesis in the Two-Higgs-doublet Model }

\author{María Olalla Olea-Romacho }
\email{mariaolalla.olearomacho@phys.ens.fr}
\affiliation{Laboratoire de Physique de l'École Normale Supérieure, ENS, Université PSL, CNRS, Sorbonne Université, Université Paris Cité, F-75005 Paris, France}

\begin{abstract}
$\gamma$-ray emission of blazars infer the presence of large-scale magnetic fields in the intergalactic medium, but their origin remains a mystery. Using recent data from MAGIC, H.E.S.S.~and $\textit{Fermi}$-LAT, we investigate whether the large-scale magnetic fields in the intergalactic medium could have been generated by a first-order electroweak phase transition in the two-Higgs-doublet model (2HDM). We study
two representative scenarios where we vary the initial conditions of the magnetic field and the
plasma, assuming either a primordial magnetic field with
maximal magnetic helicity or a primordial magnetic field with negligible magnetic helicity in a plasma
with kinetic helicity. Applying the constraints derived from MAGIC, H.E.S.S.~and $\textit{Fermi}$-LAT data, we demonstrate that a first-order electroweak phase transition within the 2HDM may account for the intergalactic magnetic fields in the case of the strongest transitions. We show that this parameter space also predicts strong gravitational wave signals in the reach of space-based detectors such as LISA, providing a striking multi-messenger signal of the 2HDM.
\end{abstract}

\maketitle

\section{Introduction}

Observations of blazars 
indirectly infer the presence of coherent magnetic fields in the intergalactic medium~\cite{MAGIC:2022piy, HESS:2023zwb, Neronov:2010gir}. The origin of these intergalactic magnetic fields (IGMF) is a longstanding puzzle, with two main scenarios proposed for their genesis: astrophysical and cosmological. Weak initial magnetic fields stem from localized effects within astrophysical objects (i.e., such as the Biermann battery mechanism~\cite{PhysRev.82.863}) that are subsequently amplified through dynamo effects~\cite{AlvesBatista:2021sln} are an example of astrophysical phenomena that can create such long-correlated magnetic fields. Such mechanisms face challenges when it comes to explaining the magnetic fields in cosmic voids with a large volume filling factor~\cite{Dolag:2010ni}.
This makes potential sources originating from early universe processes especially compelling. Primordial magnetogenesis has been addressed in the context of inflation~\cite{Turner:1987bw, Ratra:1991bn, Martin:2007ue}, post-inflationary reheating~\cite{Kobayashi:2014sga}, the electroweak (EW) phase transition~\cite{Vachaspati:1991nm, Ellis:2019tjf} or the QCD phase transition~\cite{Sigl:1996dm, Tevzadze:2012kk}. Notably, the idea of magnetogenesis during a first-order electroweak phase transition (FOEWPT) was first introduced in Ref.~\cite{Vachaspati:1991nm}. In this context, magnetic
fields are generated through EW sphaleron decays~\cite{Vachaspati:2001nb, Copi:2008he}. Since the bubbles nucleated during the EW phase transition expand, collide and merge, they stir the primordial plasma at high Reynolds
number, and the magnetic fields enter a regime of magnetohydrodynamic (MHD) turbulence~\cite{Witten:1984rs, Hogan:1986qda, Kamionkowski:1993fg, Brandenburg:1996fc, Christensson:2000sp, Kahniashvili:2010gp, Brandenburg:2017neh}. Moreover,  baryon number violating processes can produce magnetic fields with non-vanishing net helicity due to changes of the Chern-Simons number~\cite{Vachaspati:2001nb, Copi:2008he, Chu:2011tx}. Magnetic helicity has a significant impact on the evolution of magnetic turbulence over time, particularly by instigating an \textit{inverse cascade} of magnetic energy, and transferring the energy from smaller to larger length scales~\cite{Kahniashvili:2012uj, Copi:2008he}, and, thus, forming coherent magnetic structures at scales much larger than the ones where the energy was initially injected. The net magnetic helicity could be significant when there is substantial violation of the $\mathcal{C}$ and $\mathcal{CP}$ symmetries~\cite{Forbes:2000gr, Brandenburg:2017neh}, as required by the Sakharov conditions to explain the matter-antimatter asymmetry of the universe~\cite{Sakharov:1967dj}. Therefore, the observed diffuse gamma ray sky may also
hold information about the cosmological helical magnetic
field and $\mathcal{CP}$ violation in the early universe~\cite{Tashiro:2013ita, Chen:2014qva}. Furthermore, the violation of these symmetries might also affect the motion of the plasma, resulting in non-zero helicity of the velocity field~\cite{Brandenburg:2017neh}. The presence of  kinetic helicity in the plasma can ultimately source inverse cascades even in the context of non-helical magnetic fields~\cite{Brandenburg:2017rnt}. In this letter, we will study these two representative scenarios where we vary the initial conditions of the plasma and the magnetic
field generated during a FOEWPT, assuming either a primordial magnetic field with maximal magnetic helicity or a primordial magnetic field with negligible magnetic net helicity in a plasma with initial kinetic helicity.

Extended Higgs sectors have the potential to explain the observed baryon asymmetry of the universe through EW baryogenesis~\cite{Kuzmin:1985mm}. The simplest constructions to achieve EW baryogenesis extends the Standard Model (SM) scalar sector by a second Higgs doublet~\cite{Lee:1973iz}, referred to as the Two-Higgs-Doublet Model (2HDM). The FOEWPT in the 2HDM has been extensively studied in the literature~\cite{Dorsch:2013wja, Dorsch:2014qja, Dorsch:2016nrg,Basler:2016obg, Goncalves:2021egx, Biekotter:2022kgf, Biekotter:2023eil, Goncalves:2023svb}, with many of these analyses taking a multi-faceted approach to the topic, combining collider signatures of this scenario with prospects for detecting a stochastic primordial gravitational wave (GW) background at future space-based GW observatories, such as the Laser Interferometer Space Antenna (LISA)~\cite{Caprini:2019egz,Auclair:2022lcg}. This study contributes to this multi-messenger investigation of the EW phase transition in the 2HDM by presenting predictions for the generation of primordial magnetic fields during the EW phase transition, and comparing them to the most recent constraints on the strength of the IGMF presented by the \textit{Fermi}-LAT and H.E.S.S. collaborations~\cite{HESS:2023zwb} and the \textit{Fermi}-LAT and MAGIC collaborations~\cite{MAGIC:2022piy}.

This paper is structured as follows: in Sec.~\ref{sec:2HDM}, we provide a brief overview of the 
$\mathcal{CP}$-conserving 2HDM, together with the experimental and theoretical constraints that shape its physically allowed parameter space. We also present a 2HDM scenario that allows for a strong FOEWPT. In Sec.~\ref{sec:ThermalHistory}, we summarize the basis for the description of strong FOEWPTs and the computation of the GW spectrum. In Sec.~\ref{sec:PMF}, we outline how to compute the magnetic field spectrum today, depending on the two qualitatively different scenarios for the decay of MHD turbulence. In Sec.~\ref{sec:ResANDDis}, we discuss the results, and we conclude in Sec.~\ref{sec:Conclu}.

\section{Theoretical background}
\subsection{The two-Higgs-doublet model}
\label{sec:2HDM}
In this section, we provide a concise overview of the $\mathcal{CP}$-conserving 2HDM which consists of two Higgs doublets, $\Phi_{1}$ and $\Phi_{2}$, each bearing a hypercharge of $1/2$.\footnote{See e.g.~Ref.~\cite{Branco:2011iw} for a complete review of the two-Higgs-doublet model.} We impose a softly broken $\mathbb{Z}_{2}$ symmetry~\cite{Fayet:1974fj, Fayet:1974pd}, $\Phi_{1}\rightarrow\Phi_{1},\,\Phi_{2}\rightarrow-\Phi_{2}$, that prevents the existence of flavor changing neutral currents at tree-level when extended to the Yukawa sector. The tree-level scalar potential is given by
\begin{align}
\label{eq:HiggsPotential}
& V_{\text{tree}}(\Phi_1,\Phi_2) =   \\
&= m_{11}^2\,\Pdd + m_{22}^2\,\Puu - m_{12}^2\left(\Pdu + \Pud\right) \nonumber\\
&\hphantom{=} + \frac{1}{2}\lambda_1 (\Pdd)^2 + \frac{1}{2}\lambda_2 (\Puu)^2  + \lambda_3 (\Pdd)(\Puu) \nonumber        \\
&\hphantom{=} + \lambda_4 (\Pdu)(\Pud) + \frac{1}{2}\lambda_5 \left((\Pdu)^2 + (\Pud)^2\right),\nonumber 
\label{eq:HiggsPotential}
\end{align}
where all parameters are real as a result of requiring $\mathcal{CP}$-conservation. The fields vacuum expectation values (vevs) are $\langle \Phi_1 \rangle = v_1$ and  $\langle \Phi_2 \rangle = v_2$, with $v_1^2 + v_2^2 \equiv v^2 \simeq 246\GeV$ and $v_2/v_1 \equiv \tan\beta \equiv t_{\beta}$. After spontaneous symmetry breaking, the $\mathcal{CP}$-conserving 2HDM gives rise to five physical
mass eigenstates in the scalar sector: two $\mathcal{CP}$-even neutral scalars $h$ and $H$, one $\mathcal{CP}$-odd
neutral pseudoscalar $A$ and a pair of charged states $H^{\pm}$. The rotation from the gauge basis to the mass basis involves two
mixing matrices with the mixing angles $\alpha$ and $\beta$ for the $\mathcal{CP}$-even and the $\mathcal{CP}$-odd/charged sector,
respectively.

The state $h$ is conventionally chosen
as the lightest $\mathcal{CP}$-even scalar and, for the remainder of the discussion, plays the role of the discovered
Higgs boson~\cite{Aad:2012tfa, Chatrchyan:2012ufa} $h_{125}$ at $m_{h}=125 \GeV$. 
Furthermore, we adopt the so-called \textit{alignment limit}~\cite{Gunion:2002zf}, 
\begin{equation}
\mathrm{cos}(\beta-\alpha) = 0 \, ,
\end{equation}
where the lightest Higgs boson $h$ with mass $m_h = 125 \GeV$ has the SM-Higgs boson couplings at tree-level.

As the two fields $\Phi_{1}$ and $\Phi_{2}$ transform differently under the $\mathbb{Z}_{2}$
symmetry, they cannot be coupled both to the same SM fermions, which leads to four 2HDM configurations/types that avoid flavor changing neutral currents
at tree-level, characterised by the $\mathbb{Z}_{2}$ charge assignment of the fermion fields in the Yukawa sector. Therefore, to generically describe the 2HDM parameter space we must specify the Yukawa type, along with the  following set of independent parameters:
\begin{equation}
t_{\beta} \ , \; m_{12}^{2} \ ,
\; v \ , \; \cos (\beta - \alpha) \ ,
\; m_{h} \ , \;m_{H} \ , \; m_{A} \ ,
\; m_{H^{\pm}}.
\label{input_parameters}
\end{equation}
After considering both theoretical and experimental constraints and demanding a strongly FOEWPT, we will focus on a benchmark scenario with just three free parameters: $m_{A}$, $m_{H}$, and $t_{\beta}$ (see below).

\subsubsection{Constraints}
\label{sec:constraints}
Employing the publicly available \texttt{thdmTools} code~\cite{Biekotter:2023eil}, we impose a set of constraints on the parameter space of the 2HDM. These constraints encompass both theoretical and experimental aspects, including:
\begin{itemize}
    \item Perturbative unitarity.
    \item Vacuum stability.
    \item Direct searches for BSM scalars using \texttt{HiggsTools}~\cite{Bahl:2022igd}.
\end{itemize}
\subsubsection{A benchmark scenario for a FOEWPT in the 2HDM}
\label{sec:scenario}

We explore a 2HDM parameter region that is especially compelling for the realization of a strong FOEWPT. In this scenario, the $\mathcal{CP}$-odd scalar $A$ and the charged scalars $H^{\pm}$ are enforced to be mass-degenerate, $m_{A} = m_{H^{\pm}}$, a choice motivated by EW precision data.
Moreover, we assume the alignment limit $\mathrm{cos}(\beta-\alpha) = 0$, which is favored by the measurements of 
the signal rates of $h_{125}$~\cite{ATLAS:2022vkf, CMS:2022dwd}.
The $\mathbb{Z}_{2}$-breaking mass scale $M^2 = m_{12}^2/(s_{\beta} c_{\beta})$ is set equal to the mass of the heavy $\mathcal{CP}$-even scalar $H$, i.e.~$M=m_H$.  These conditions allow for large $m_{A}-m_{H}$ mass splittings, driven by the requirement for large quartic couplings in the 2HDM potential,
which enable a FOEWPT~\cite{Dorsch:2013wja, Dorsch:2014qja, Biekotter:2022kgf}  while agreeing with the LHC data on the 125 GeV Higgs boson, the measurements of EW precision observables, and other theoretical constraints.

Furthermore, here we will concentrate on the Yukawa type II. Given that all four Yukawa types share identical couplings between the neutral scalars and the top quark, we expect only minor differences among these types concerning the phase transition dynamics.

Taking into account the above, the remaining free parameters are $m_{H}$, $m_{A} = m_{H^{\pm}}$, and $t_\beta$.

\subsection{Thermal history analysis}
\label{sec:ThermalHistory}
To investigate the dynamics of the EW phase transition in the 2HDM, we will follow the formalism of the temperature-dependent effective potential (see e.g.~Ref.~\cite{Quiros:1999jp} for a review). The full effective potential is given by
\begin{align}
    V_{\rm eff}(\phi_i,T) &= V_{\rm tree}(\phi_i) + V_{\rm CW}(\phi_i) +
        V_{\rm CT}(\phi_i) \\ \nonumber
        &+ V_{\rm T}(\phi_i,T) + V_{\rm daisy}(\phi_i,T).
    \label{eq:fullpotential}
\end{align}
The temperature-independent part consists of the tree-level scalar potential $V_{\rm tree}$, given by Eq.~\eqref{eq:HiggsPotential}, and the one-loop Coleman-Weinberg potential $V_{\rm CW}$~\cite{Coleman:1973jx}. We also include a set of UV-finite counterterms $V_{\rm CT}$ which enforces the zero-temperature loop-corrected vacuum expectation values, scalar masses and mixing angles to be equal to their tree-level values~\cite{Basler:2016obg}. The temperature-dependent part comprises the one-loop thermal corrections $V_T$ to the scalar potential~\cite{Dolan:1973qd}, and the resummation of the daisy-diagrams $V_{\rm daisy}$ in the Arnold-Espinosa scheme~\cite{Arnold:1992rz}.

 It is important to note that 
 this approach carries significant theoretical uncertainties~\cite{Curtin:2016urg, Croon:2020cgk,Gould:2021oba, Gould:2023ovu}, and therefore the parameter space explored in the subsequent analysis should be considered only as indicative of a first-order phase transition.

In a first-order phase transition, the scalar fields involved in the transition undergo a discontinuous change from a high-temperature symmetric phase (the false vacuum) to a low-temperature broken phase (the true vacuum) as the universe cools down. In this scenario, bubbles of true vacuum nucleate and expand in the background of the false vacuum, eventually converting all space into the true vacuum. The false and true vacuums are characterized by the values of the scalar fields expectation value, which are zero and nonzero, respectively. The onset of the transition depends on the nucleation rate per unit time and volume~\cite{Coleman:1977py,Callan:1977pt,Linde:1980tt,Linde:1981zj}
 \begin{equation}
\Gamma(T)=A(T)\,e^{-S(T)/T},
\label{eqfuncdeter}
\end{equation}
where $S(T)$ is the three-dimensional Euclidean action of the $O(3)$ symmetric bounce solution. By requiring that on average one bubble is nucleated per horizon
volume, we define the nucleation temperature~\cite{Auclair:2022lcg}
\begin{equation}
S(T_n) / T_n \sim 140 \,.
\label{cond_trans}
\end{equation}
In the 2HDM, the temperature at which bubbles of the new phase start to form (nucleation) and the temperature at which they fill the whole space (percolation) are very similar~\cite{Biekotter:2023eil}. Therefore, we use the same symbol $T_{*}$ to refer to both temperatures, and we call it the transition temperature.
We track the co-existing minima of the effective potential as a function of the temperature and compute the nucleation rate by using \texttt{cosmoTransitions}~\cite{Wainwright:2011kj}.

Apart from the transition temperature $T_{*}$, the first-order cosmological phase transition is characterized by three other quantities: (i) the strength at the transition temperature $\alpha$ (normalized to the
energy density in the plasma at the time of nucleation), (ii) the inverse duration of the transition (in Hubble units) $\beta/H$, and (iii) the bubble wall velocity in the rest frame of the fluid (away from the bubble) $v_w$. The strength of the transition is given by
the ratio of the difference of the trace of the
energy-momentum tensor between the two phases (false and true vacua) to the background radiation energy density, i.e.~\cite{Caprini:2019egz,Auclair:2022lcg}
\begin{equation}
\alpha =
\frac{1}{\rho_{R}}
\left(
\Delta V(T_{*})-\left.\left(
\frac{T}{4}\frac{\partial
\Delta V(T)}{\partial T}\right)\right|_{T{*}}
\right) \ ,
\label{eqalpha}
\end{equation}
where $\Delta V(T)=V_{f}-V_{t}$, with $V_f \equiv V_{\rm eff}(\phi_f)$ and $V_t \equiv V_{\rm eff}(\phi_t)$ being the values of the potential in the false and true vacuum.
The inverse duration of the transition in Hubble units $\beta/H$ can be defined as
\begin{equation}
\frac{\beta}{H} =
T_{*}\left.\left(\frac{d}{dT}\frac{S(T)}{T}\right)\right|_{T_{*}}.
\label{beta}
\end{equation}
Finally, the the fourth quantity that characterizes a cosmological first-order phase transition is the bubble wall velocity $v_{w}$. Except for the case of ultrarelativistic bubbles~\cite{Bodeker:2009qy,Bodeker:2017cim}, computing $v_{w}$ is a challenging task that requires solving a coupled system of Boltzmann and scalar field equations in a model-dependent approach~(see e.g.~Refs.~\cite{Moore:1995si,Konstandin:2014zta,Kozaczuk:2015owa,Dorsch:2018pat,BarrosoMancha:2020fay,Laurent:2020gpg,Dorsch:2021nje,Ai:2021kak,Lewicki:2021pgr,Laurent:2022jrs}. Recent results suggest that phase transition bubbles tend to expand with either $v_{w} \approx c_s$ ($c_s$ being the speed of sound of the plasma)\footnote{For a relativistic perfect fluid, $c_s = 1/\sqrt{3} \simeq 0.577$.} or $v_{w} \to 1$~\cite{Laurent:2020gpg,Ai:2023see} (see also Ref.~\cite{Dorsch:2018pat} for more discussion on bubble wall velocity estimates in BSM theories). In this work we fix $v_w = 0.6$ as an optimistic case.\footnote{See Ref.~\cite{Jiang:2022btc} for estimates of the bubble wall velocity in a similar model.}

\subsubsection{Electroweak baryogenesis}

The baryon asymmetry of the universe can arise dynamically from EW baryogenesis~\cite{Kuzmin:1985mm}. In this scenario, the FOEWPT provides the out-of-equilibrium condition, one of the requirements for successful baryogenesis according to the Sakharov conditions~\cite{Sakharov:1967dj}. 
To avoid the washout of the baryon asymmetry after the phase transition, the following criterion has been often used in the literature~\cite{Dimopoulos:1978kv}
\begin{equation}
\xi_n \equiv v(T_n)/T_n \gtrsim 1 \, .
\label{eq:strongFOPT}
\end{equation}
where $v(T_n) = \sqrt{v_1(T_n)^2+v_2(T_n)^2}$.
This condition also characterizes a strong FOEWPT.
Besides this requirement, the SM needs additional sources of $\mathcal{CP}$-violation to generate the matter-antimatter asymmetry. We will not perform a detailed study of EW baryogenesis here (see Ref.~\cite{Goncalves:2023svb} for a recent study), but we will focus only on one of its essential elements, the FOEWPT, assuming that the properties of the transition are not significantly affected by the inclusion of $\mathcal{CP}$-violation. 

\subsubsection{Gravitational wave relics}
Cosmological phase transitions that are first-order may produce a stochastic GW background~\cite{Witten:1984rs, Hogan:1986qda} that can be probed by future GW interferometers. For a FOEWPT, the GW spectrum peaks around mHz frequencies, which matches the optimal sensitivity range of the space-based interferometer LISA~\cite{LISA:2017pwj, LISACosmologyWorkingGroup:2022jok}. In the 2HDM, the GW spectrum is mainly sourced by the plasma motions after the bubble collisions, in the form of sound waves and MHD turbulence, rather than by the bubble wall collisions themselves~\cite{Dorsch:2016nrg}. We use numerical power-law fits to the results of hydrodynamical simulations of the thermal plasma, which modeled the GW stochastic background as a function of the four parameters defined above. The formulas for the GW spectral
shapes, amplitudes and peak frequencies are taken from Ref.~\cite{Biekotter:2022kgf}, which follows Refs.~\cite{Caprini:2019egz,Auclair:2022jod}.
The detectability of a stochastic GW signal at a GW observatory
depends on the signal-to-noise
ratio~(SNR),
which is given by
\begin{equation}
{\rm SNR} = \sqrt{\mathcal{T} \int^{+ \infty}_{- \infty} {\rm d}f \left[\frac{h^2\Omega_{\rm GW}(f)}{h^2\Omega_{\rm Sens}(f)}\right]^2},
\label{eq:SNR}
\end{equation}
where $\mathcal{T}$ is the duration of the experiment, $h^2 \Omega_{\rm Sens}$ is the nominal sensitivity of
the detector, according to the mission requirements~\cite{LISAmissionreq},
and $h^2 \Omega_{\rm GW} = h^2 \Omega_{\rm sw} + h^2 \Omega_{\rm turb}$ is the spectral shape of the GW signal.
We focus on the GW detectability with LISA, for which
we assume an operation time of $\mathcal{T}$ = 7 years, and consider a GW signal to be detectable if
SNR~$>1$.

\subsection{Primordial magnetic fields}
\label{sec:PMF}

\subsubsection{Magnetic field spectrum today}
The decay of MHD turbulence is influenced significantly by initial conditions and physical processes which are presently uncertain. However, some predictions can be derived based on different assumptions~\cite{Armua:2022rvx}.
One of the factors that affects the decay of MHD turbulence is the magnetic helicity of the initial seed field~\cite{Banerjee:2004df}, which is a measure of the twist and linkage of the magnetic field lines. The average helicity is defined as $\langle \mathbf{A} \cdot \mathbf{B} \rangle$, where $\mathbf{B}=\nabla \times \mathbf{A}$.
In a highly conductive medium, the magnetic helicity is approximately conserved. This means that a maximally helical field has to increase its correlation length as it loses magnetic energy, which results in an \textit{inverse cascade} of magnetic energy where the energy is transferred from smaller scales to larger scales, forming coherent magnetic structures at scales much larger than the ones where the energy was initially injected. Therefore, this phenomenon could have played a crucial role in the survival and evolution of primordial magnetic fields, which are correlated at very large length scales today.

For maximally helical magnetic fields, the MHD turbulence, $B$, decays as a power law in conformal time, $t$, with
the following scaling relations for the magnetic energy
and correlation length during the radiation-dominated
epoch~\cite{PhysRevLett.83.2195}:
\begin{equation}
B \sim t^{-1/3} \, \, \mathrm{and} \, \, \lambda \sim t^{2/3}.
\end{equation}

Numerical simulations have shown that non-helical magnetic fields, which have zero or negligible net helicity, can also undergo an inverse cascade of magnetic energy in the presence of a plasma with initial kinetic helicity~\cite{Brandenburg:2017rnt}.\footnote{This has been contested recently in Ref.~\cite{Armua:2022rvx}, where it was found a weaker inverse transfer of magnetic energy for non-helical fields, compared
to previous studies in the literature.}  In this case, the magnetic energy and
correlation length, $\lambda$, scale as
\begin{equation}
    B \sim t^{-1/2} \,\, \, \mathrm{and} \,\, \, \lambda \sim t^{1/2} \, ,
\end{equation}
respectively, during the radiation-dominated epoch. These scaling laws are valid until the epoch of matter-domination, when the scale factor grows linearly with conformal time, i.e.~$a \sim t$. After recombination, the magnetic field redshifts like radiation, i.e.~$B \sim a^{-2}$.
To express these two scenarios in a compact way, we define
the parameters
\begin{equation}
q_b = \frac{2}{b+3} \, \, \, \, \mathrm{and} \, \, \, \,  p_b = \frac{2}{b+3}(b+1) \, ,
\end{equation}
for the power-laws
\begin{equation}
    B \sim t^{- p_b / 2} \, \,\, \, \mathrm{and} \,\, \, \, \lambda \sim t^{q_b} \, ,
\end{equation}
where the cases $b=0$ and $b=1$ correspond to the maximally helical and non-helical scenarios described above, respectively.

In the 2HDM scenario, where the expanding bubbles do not enter the run-away regime~\cite{Dorsch:2016nrg}, the magnetic field generated by the bubble collisions can be safely ignored. Thus, the magnetic field energy density 
at percolation temperature can be estimated by~\cite{Ellis:2019tjf, RoperPol:2023bqa}
\begin{equation}
\rho_{B,*} = \varepsilon_{\rm turb} K  \rho_{*} = 0.1\frac{ \kappa \alpha}{1+ \alpha} \rho_{*}.
\end{equation}
Here, $\rho_{*}=3M_p^2H_{*}^2$ is the total energy density at the percolation temperature and $K=\kappa \alpha/(1+ \alpha)$ is the ratio of the kinetic energy density of the sound waves to the total energy density. $\kappa$ is the fraction of the released vacuum energy that is converted into the kinetic energy density of the
plasma~\cite{Espinosa:2010hh} and it was obtained following Ref.~\cite{Espinosa:2010hh}. 
Additionally, $\varepsilon_{\rm turb}$ denotes the fraction of sound wave kinetic energy density expended on magnetic field generation, with numerical simulations suggesting $\varepsilon_{\rm turb} \approx 0.1$~\cite{Kahniashvili:2009qi,Durrer:2013pga, Brandenburg:2017neh}.

The magnetic field spectrum today can be computed as~\cite{Ellis:2019tjf}
\begin{align}
\label{eq:spectrum}
&B_0(\lambda) \equiv B(\lambda,t_0) = \\ &\left(\frac{a_*}{a_{\rm rec}}\right)^{p_b/2} \left(\frac{a_*}{a_0}\right)^2 \sqrt{\frac{17}{10}\,\rho_{B,*}} \,
\begin{cases}
\left(\frac{\lambda}{\lambda_0}\right)^{-5/2}  & \mbox{ for }~~ \lambda\geq\lambda_0 \, \\    \left(\frac{\lambda}{\lambda_0}\right)^{1/3} & \mbox{ for }~~ \lambda < \lambda_0 \,, \nonumber
\end{cases}
\end{align}
which assumes a power-law spectrum for the magnetic field strength, with a spectral index of $n=2$ at large scales. 
Here $\lambda_0$ denotes the field coherence scale redshifted to today~\cite{Ellis:2019tjf}
\begin{equation}
\lambda_0 \equiv \lambda_B(t_0) = \left(\frac{a_{\rm rec}}{a_*}\right)^{q_b} \left(\frac{a_0}{a_*}\right) \lambda_*, \,
\end{equation}
where the initial correlation length $\lambda_{*}$ is given by the bubble size at percolation $R_{*}$
\begin{equation}
\lambda_{*} = R_{*} = \frac{(8 \pi)^{1/3}}{H_{*}} \left( \frac{\beta}{H} \right)^{-1}v_w.
\end{equation}
The redshift factors are computed as
\begin{align}
&\frac{a_{*}}{a_0} = 8 \cdot 10^{-14} \left( \frac{100}{g_{*}} \right)^{1/3} \left( \frac{1 \GeV}{T_{*}} \right) \, , \\
&\frac{a_{*}}{a_{\rm rec}} = 8 \cdot 10^{-11} \left( \frac{100}{g_{*}} \right)^{1/3} \left( \frac{1 \GeV}{T_{*}} \right) \, ,
\end{align}
where $g_{*}$ corresponds to the total number of relativistic degrees of freedom in entropy at the transition temperature $T_{*}$.

\subsubsection{Experimental constraints from blazar emissions}

Active galactic nuclei jets that are approximately
pointed in our direction are called blazars, and they are sources of TeV $\gamma$-rays. When these $\gamma$-rays collide with photons from the extragalactic background, they give rise to $e^{+}e^{-}$ pairs. Subsequently, these pairs interact with photons from the cosmic microwave background (CMB), producing $\gamma$-rays with energies in the GeV range. This in turn changes the original spectrum of the blazars by reducing the number of TeV $\gamma$-rays and increasing the number of GeV $\gamma$-rays~\cite{Vachaspati:2016xji,Vachaspati:2020blt}. In the presence of IGMF, the trajectories of the $e^{+}e^{-}$ pairs can be deflected due to the Lorentz
force and, if the field is strong enough, the final GeV photons are no longer directed towards the observer. Non-observation of these photons has been used to set lower bounds on the strength of the IGMF. 

We considered two recent analyses of blazar observations that measure the minimum strength of the IGMF. The first one, performed by the MAGIC $\gamma$-ray observatory, sets a lower bound of $ B > 1.8 \cdot 10^{-17} \, \mathrm{G}$ for correlation lengths $\lambda \geq 0.2 \, \mathrm{Mpc}$~\cite{MAGIC:2022piy}. The second one, based on the data from the \textit{Fermi}-LAT and H.E.S.S.~collaborations, establishes a limit of $ B > 7.1 \cdot 10^{-16} \, \mathrm{G}$ for coherence lengths of $\lambda \geq 1 \, \mathrm{Mpc}$, assuming a blazar duty cycle of $t_d = 10 \, \mathrm{yrs}$~\cite{HESS:2023zwb}. The duty cycle duration is the main source of systematic uncertainty in the latter analysis, so in our conservative approach, we only use the bounds obtained for $t_d = 10 \, \mathrm{yrs}$.

\section{Results and Discussion}
\label{sec:ResANDDis}

We considered two different cases for the evolution of the MHD turbulence during the cosmic expansion. We varied the initial conditions of the magnetic field and the plasma, assuming either a primordial magnetic field with maximal magnetic helicity ($b=0$) or a primordial magnetic field with negligible magnetic helicity ($b=1$) in a plasma with kinetic helicity. We expect some fraction of magnetic helicity to be generated after the phase transition from the decay of EW sphalerons~\cite{Vachaspati:2001nb}. We can also anticipate that some helicity will be produced as the plasma vortices create twisted field lines, due to the interaction between magnetic field and plasma elements in an MHD system~\cite{Ellis:2019tjf, DieterBiskamp_2003}. It is plausible to consider that a realistic scenario may lie somewhere between the extreme case of maximal initial magnetic helicity ($b=0$) and the case with negligible magnetic helicity and non-zero kinetic helicity ($b=1$). Given a 2HDM benchmark, it is typically expected that the predictions for the value of $B_0$ and $\lambda_0$ would be larger for the case with $b=0$.

\begin{figure}[h!]
\includegraphics[width=0.47\textwidth]{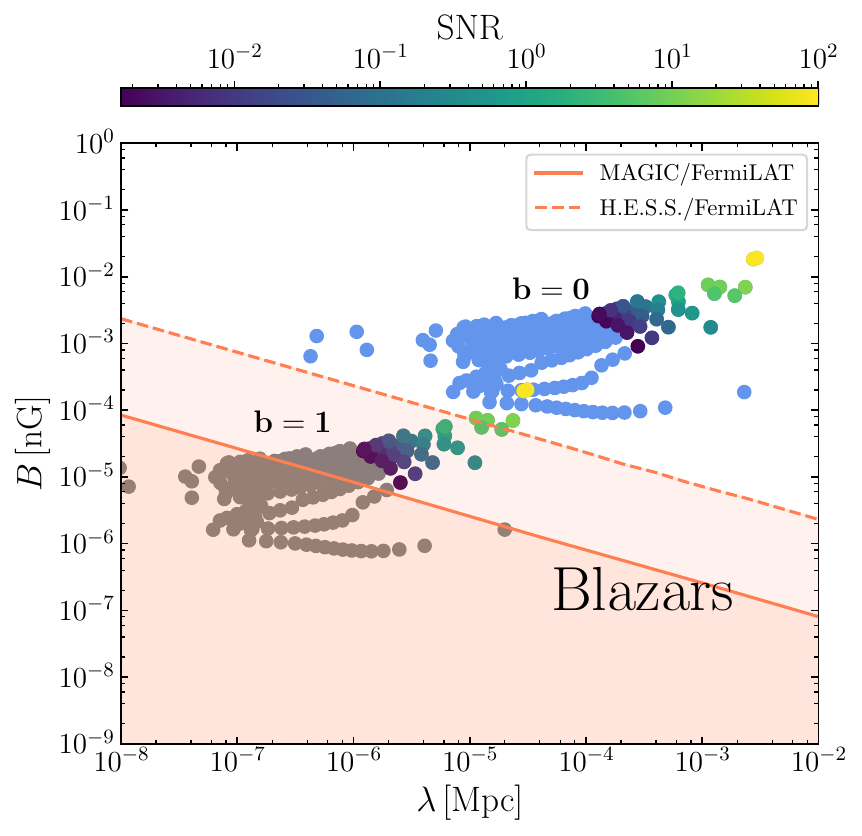}
\caption{Magnetic field strength as a function of its coherence length. The solid coral line corresponds to the lower bound on the magnetic field set by Ref.~\cite{MAGIC:2022piy}, while the dashed line corresponds to the one set by Ref.~\cite{HESS:2023zwb} for a duty cycle of $t_d = 10 \, \mathrm{yrs}$. The points indicate the values of the peak magnetic field strengths $B_0$ and the corresponding coherence lengths $\lambda_0$, for all the points of the scan in Fig.~\ref{fig:B0_mass_plane} (for $b=0$ and $b=1$). The colorbar indicates the SNR in LISA. Points for which $\mathrm{SNR}<10^{-3}$ are shown in blue ($b=0$) and in gray ($b=1$). 
}
\label{fig:SNR}
\end{figure}

In order to explore the connection between GWs at LISA and primordial magnetic fields, we show the relation between the peak value of the magnetic field $B_0$ and its coherence length $\lambda_0$ for our sample of points in Fig.~\ref{fig:SNR}. We varied the values of $m_H$ between $360 \GeV \leq m_H \leq 900 \GeV$ and $m_A$ between $560 \GeV \leq m_H \leq 1050 \GeV$. We fixed the value of $t_{\beta}$ to 3, with the rest of the parameters set as described in Sec.~\ref{sec:scenario}. We imposed all the theoretical and experimental constraints listed in Sec.~\ref{sec:constraints}.  We only considered the points that have $\xi_n \gtrsim 1$, which are preferred by EW baryogenesis. The two sets of data points correspond to the two different cases for the evolution of the MHD turbulence ($b=0$ and $b=1$). The colorbar indicates the SNR in LISA for points that satisfy $\mathrm{SNR} > 10^{-3}$. For $b=0$, the points with $\mathrm{SNR} < 10^{-3}$ are indicated in blue, whereas for $b=1$, we use gray. The solid coral line corresponds to the lower bound on the magnetic field set by Ref.~\cite{MAGIC:2022piy}, whereas the dashed line corresponds to that set by Ref.~\cite{HESS:2023zwb} for a duty cycle of $t_d = 10 \, \mathrm{yrs}$. Upper limits on the magnetic field strength derived from CMB and Big Bang Nucleosynthesis data exist for values of $B_0 \gtrsim 1 \, \mathrm{nG}$~\cite{AlvesBatista:2021sln}. 

Regardless of the choice for $b$, all benchmark points with $\mathrm{SNR} > 10^{-3}$ lie above the most conservative bound set by MAGIC/\textit{Fermi}-LAT. Moreover, data points with the highest SNR ($\mathrm{SNR}>1$), which might be detected by LISA, also have the highest peak values of the magnetic field $B_0$, and they all exceed the lower bound from blazar observations by the stricter bound set by H.E.S.S./\textit{Fermi}-LAT (for both $b=0$ and $b=1$). Therefore, regardless of the choice of $b$ and within this 2HDM framework, observable GWs are associated with strong primordial magnetic fields, consistent with blazar observations.

For $b=0$ (maximally helical fields), all benchmark points are significantly above the lower limits set by both experiments, suggesting that their observations can be accounted for by primordial magnetic fields generated within a 2HDM framework. For this case, the peak coherence length of the data set falls within the range $ 4.27 \times 10^{-7} \, \rm{Mpc} \le \lambda_0 \le 2.94 \times 10^{-3} \, \rm{Mpc}$, and the magnetic field value is bounded between $9.09 \times 10^{-5} \, \rm{nG} \le B_0 \le 1.90 \times 10^{-2}\, \rm{nG} $.

For $b=1$ (non-helical magnetic fields in a plasma
with kinetic helicity), the peak coherence length of the scan points ranges from $ 3.80 \times 10^{-9} \, \rm{Mpc} \le \lambda_0 \le 3.12 \times 10^{-5} \, \rm{Mpc}$, and the magnetic field value lies within the interval $7.69 \times 10^{-7} \, \rm{nG} \le B_0 \le 2.02 \times 10^{-4} \, \rm{nG}$.
In this case, if we focus on the higher lower limit by \textit{Fermi}-LAT and H.E.S.S., only the points with the strongest GW signals could account for the origin of the primordial magnetic fields.

For both scenarios ($b=0$ and $b=1$), this work exemplifies the need for continued improvement in the analysis of magnetic fields via blazar observations to exclude primordial seed fields sourced by the EW phase transition in the 2HDM. Conversely, GW interferometry can serve as 
an additional probe of a 2HDM origin of primordial magnetic fields over certain coherence length scales.

\begin{figure}[h!]
\includegraphics[width=0.47\textwidth]{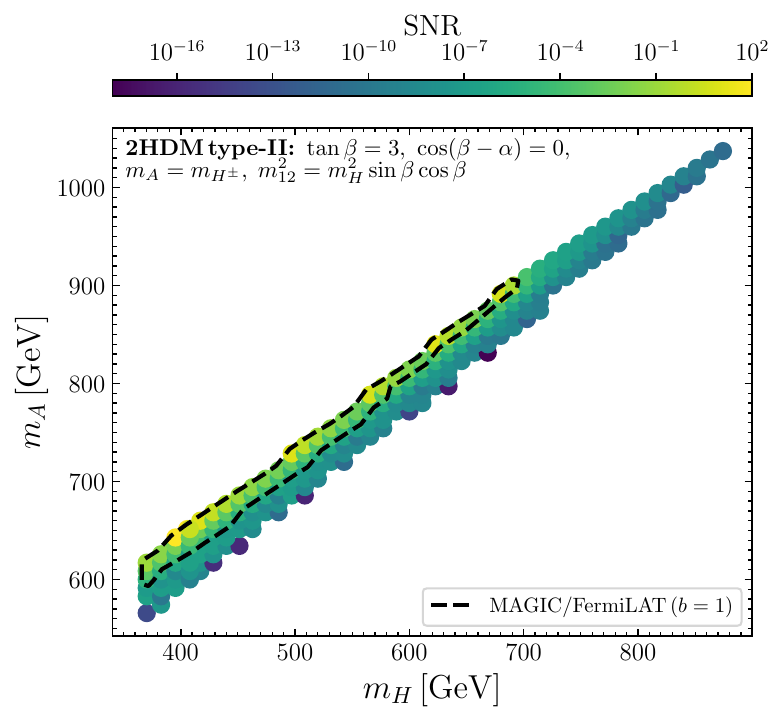}
\caption{($m_{H}$,$m_{A}$) mass plane for $t_{\beta}=3$. The colorbar indicates the SNR at LISA. The dashed black line comprises the points that exceed the lower bound on $B_0$ from MAGIC and \textit{Fermi}-LAT observations for the case of non-helical magnetic fields ($b=1$).}
\label{fig:B0_mass_plane}
\end{figure}

In Fig.~\ref{fig:B0_mass_plane}, we show the ($m_{H}$,$m_{A}$) mass plane for the aforementioned settings for the rest of the parameters of the model. 
As mentioned above, only the points with $\xi_n \gtrsim 1$ were considered, since they are favoured by EW baryogenesis. The largest value found for $\xi_n$ is 4.07. The SNR in LISA, assuming the experimental choices indicated under Eq.~(\ref{eq:SNR}) is depicted in the colorbar.

For $b=0$, all parameter points in  Fig.~\ref{fig:B0_mass_plane} produce primordial magnetic fields that could explain the observations made by both experiments.

Assuming $b=1$, only the points inside the black dashed line exceed the less stringent lower bound on $B_0$ from MAGIC and \textit{Fermi}-LAT observations. This region of parameter space also corresponds to the prediction of the strongest gravitational waves, which can be tested in future LHC runs in searches such as $gg \rightarrow A \rightarrow Z H \rightarrow \ell^+ \ell^- t \bar t$~\cite{Biekotter:2023eil}. This is another experimental probe that could reveal whether primordial magnetic fields may have a 2HDM origin or not (for relatively low values of $t_{\beta}$).

\newpage

\section{Conclusions}
\label{sec:Conclu}
We explored the possibility that the large-scale magnetic fields in the intergalactic medium could have originated from a first-order electroweak phase transition in the two-Higgs-doublet model (2HDM). We considered two scenarios with different initial conditions for the magnetic field and the plasma, assuming either a primordial magnetic field with
maximal magnetic helicity (defined by $b=0$) or a primordial magnetic field with negligible magnetic helicity in a plasma
with kinetic helicity (defined by $b=1$). We calculated the maximum strength of the magnetic field spectrum $B_0$ for a representative parameter plane (see Fig.~\ref{fig:B0_mass_plane}) that realizes a first-order electroweak phase transition in the 2HDM. We confronted these predictions with observational constraints on the intergalactic magnetic fields from MAGIC, H.E.S.S.~and $\textit{Fermi}$-LAT experiments.

We found that the strongest transitions, with a $\rm{SNR}$ for LISA larger than $10^{-3}$, can explain the $\gamma$-ray observations by $\textit{Fermi}$-LAT/MAGIC, regardless of our choice of the initial conditions for the plasma and the magnetic field (see Fig.~\ref{fig:SNR}).

$\textit{Fermi}$-LAT/MAGIC provide the most conservative bound on $B_0$ as a function of $\lambda_0$, compared to the limit obtained by $\textit{Fermi}$-LAT/H.E.S.S. Maximally helical magnetic fields, spanning scales of about $ 4.27 \times 10^{-7} \, \rm{Mpc} \le \lambda_0 \le 2.94 \times 10^{-3} \, \rm{Mpc}$, are consistent with the bounds of both experiments. For non-helical magnetic fields, only a few points in our parameter space could exceed the stricter limit by $\textit{Fermi}$-LAT/H.E.S.S.

 However, all points that predict very strong gravitational wave signals in LISA ($\rm{SNR} \geq 1$) also predict a large peak magnetic field above the $\textit{Fermi}$-LAT/H.E.S.S.~lower bound, offering a multi-messenger test of the 2HDM origin of primordial magnetic fields. Additionally, the low-$t_{\beta}$ parameter space (see Fig.~\ref{fig:B0_mass_plane}) can be explored in upcoming LHC runs. The process $gg \rightarrow A \rightarrow Z H \rightarrow \ell^+ \ell^- t \bar t$~\cite{Biekotter:2023eil} offers a collider-based approach to assess the 2HDM contribution to the genesis of intergalactic magnetic fields.

We concluded that a first-order electroweak phase transition within the 2HDM may explain (over certain coherence length scales) the  intergalactic magnetic fields inferred from blazar observations, depending on the initial conditions of the magnetic field and the plasma. We also demonstrated that this scenario could be tested by future LHC runs and LISA observations, providing a tantalizing multi-messenger signal of the 2HDM. We also emphasized the need for improved analysis of magnetic fields via blazar observations to exclude or confirm such model predictions, i.e.~primordial magnetic fields sourced by the electroweak phase transition in the 2HDM.

\vspace{10mm}

\section*{Acknowledgments}
I am thankful to Shyam Balaji, Iason Baldes, Thomas Biekötter, Alberto Roper Pol and Ramkishor Sharma for useful discussions. My work is supported by the European Union's
Horizon 2020 research and innovation programme under grant agreement No 101002846, ERC CoG ``CosmoChart''.

\bibliography{references}
\bibliographystyle{apsrev4-1}

\end{document}